\documentclass[]{jfm}
\usepackage{graphicx}
\usepackage{newtxtext}
\usepackage{newtxmath}
\usepackage{natbib}
\usepackage{hyperref}
\hypersetup{
    colorlinks = true,
    urlcolor   = blue,
    citecolor  = black,
}

\newcommand{\RomanNumeralCaps}[1]
\linenumbers

\usepackage{layouts} 
\title{Navier-Stokes-driven analysis of mean and fluctuating wall
  shear stress in turbulent channel flow}

\author{Le Yin\aff{1}
  \corresp{\email{le.yin@centralelille.fr}},
  Yongyun Hwang\aff{2}
  \and John Christos Vassilicos\aff{1}
  \corresp{\email{john-christos.vassilicos@cnrs.fr}}}

\affiliation{\aff{1}UMR 9014 - LMFL - Laboratoire de Mécanique des Fluides de Lille - Kampé de Fériet, Univ. Lille, CNRS, ONERA, Arts et Metiers Institute of Technology, Centrale Lille, F-59000 Lille, France
\aff{2}Department of Aeronautics, Imperial College London, South Kensington, London SW7 2AZ, UK}

\begin{document}
\maketitle

\begin{abstract}
We propose a Navier-Stokes-driven analysis of the mean and fluctuating
wall shear stress (WSS) applied to turbulent channel flow data from
direct numerical simulations at friction Reynolds numbers up to
$Re_\tau\approx 2000$.  Starting from the streamwise momentum
equation, we derive exact integral equations that relate the square
plane-average and the square fluctuating WSS to wall-normal integrals
of terms combining shear with acceleration, shear with
pressure-gradient, and shear with viscous diffusion.  The square
plane-average WSS can be well approximated by the product of
plane-average shear and plane-average acceleration integrated over the
buffer layer
with corrections from the mean pressure gradient which diminish as the
reciprocal of the Reynolds number. The square fluctuating WSS is
similarly well approximated by the shear-acceleration and
shear-pressure-gradient covariances integrated over the buffer layer,
but the latter
increases in magnitude with Reynolds number and is therefore not
negligible. The acceleration fluctuations around the plane-average
acceleration consist of a local Eulerian fluctuating acceleration, an
advective acceleration and a term which gives rise a turbulence
production contribution to the shear-acceleration covariance. By
Taylor's frozen turbulence hypothesis the Eulerian acceleration and
the streamwise mean advection part of the advective acceleration
cancel each other. The shear-acceleration covariance is characterised
by a near-wall peak which results from turbulence production and, more
specifically, sweeps.

\end{abstract}


%%%%%%%%%%%%%%%%%%%%%%%%%%%%%%%%%%%%%%%%%%%%%%%%%%%%%%%%%%%%%%%%%%%%%%
\section{Introduction}\label{sec:intro}
Real world turbulent flows are often, if not always, bounded by walls.
The flow may be considered external in the case of a turbulent
boundary layers (TBL) or internal in the case of a turbulent channel
flow (TCF) or turbulent pipe flow (TPF). The existence of a no-slip
boundary, together with fluid viscosity, gives rise to shear stresses
close to the wall, which relaxe further away from the wall. Right at
the wall, the stresses are naturally named wall shear stress. The
surface integral of shear stresses at the wall in the streamwise
direction constitutes a parallel force acting on the surface by the
fluid, i.e. skin friction drag. The flow above the wall is often
turbulent, and since turbulent flows are dynamic, i.e. fluctuating in
space and time, the instantaneous and local-in-space wall shear
stresses must also be fluctuating. The relationship between the wall
shear stress and above fluid motion is best described by the
Navier-Stokes equations. Understanding the dynamical process of wall
shear stress necessitates studies of both the mean and the
fluctuations around the mean, and is perhaps best undertaken through
an analysis based on the Navier-Stokes equations.

\subsection{Mean wall shear stress and skin friction identities}
\label{sec:fik}
Many attempts have been made to relate the mean wall shear stress with
physical processes above the wall by integrating the Navier-Stokes
equations under certain assumptions. The employment of integration is
a natural way to obtain relations of skin friction from the
Navier-Stokes equations, since the skin friction comes out of the
integrated wall-normal viscous diffusive term. \citet{vonkarman_1921}
first integrated the streamwise boundary layer-approximated
Navier-Stokes equation in the wall-normal direction from the wall to
infinity and related the mean skin friction directly to displacement
thickness and momentum thickness.  \citet{ligrani_moffat_1986} adopted
a similar strategy by integrating up to a finite distance from the
wall. They obtained a momentum balance equation which was then used to
experimentally estimate the skin friction from velocity and Reynolds
shear stress measurement above the wall.

Integral relations have attracted more interest in the past two
decades. \citet{fik_2002} derived a relation (FIK identity) by
integrating the streamwise Reynolds-averaged Navier-Stokes equation
three times in the wall normal direction. They obtained a relation
(\ref{eq:fik}) (in standard notation which, for economy of space, we
do not define in this qualitatively descriptive intoduction) which
clearly distinguishes laminar and turbulent contributions to the mean
skin friction coefficient $C_f$, plus an non-stationary
non-homogeneous term,
\begin{equation}
C_f = 
\underbrace{\frac{12}{Re_b}}_{\mathrm{Laminar}} 
+ \underbrace{\int^\delta_0 -(1-y) \langle u'v'\rangle ~\mathrm{d}y }_{\mathrm{weighted~Reynolds~stress}} 
~\underbrace{ -12 \int^\delta_0 (1-y)^2 \left( I_x'' + \frac{\partial p''}{\partial x} 
+ \frac{\partial \overline{u}}{\partial t}\right)~\mathrm{d}y.}_\mathrm{Non-stationary~non-homogeneous~term}
\label{eq:fik}
\end{equation}
\citet{renard_deck_2016} derived another relation (RD identity) of
skin friction by integrating the mean energy equation once in the
wall-normal direction in an absolute frame of reference moving with
either the bulk flow velocity in internal flow, or the free stream
velocity in external flow.  While the RD identity (\ref{eq:rd}) does
not separate the laminar from the turbulent contributions as the
effect of turbulence is implied in the mean shear profile which enters
the mean dissipation, it describes the contributions to the skin
friction through the mean dissipation and turbulent energy production
plus the additional non-homogeneous term. In the case of TCF, the
relation can be written as,
\begin{equation}
C_f = 
\underbrace{\frac{2}{U_b^3} \int_0^\delta \nu \left( \frac{\partial \langle u \rangle}{\partial y}\right)^2 ~\mathrm{d}y}_\mathrm{Mean~dissipation} +
\underbrace{\frac{2}{U_b^3} \int_0^\delta -\langle u'v' \rangle  \frac{\partial \langle u \rangle}{\partial y} ~\mathrm{d}y}_\mathrm{Turbulence~production}  +
\underbrace{\frac{2}{U_b^3} \int_0^\delta(\langle u \rangle - U_b)\frac{\partial}{\partial y}\left(\frac{\tau}{\rho}\right)~\mathrm{d}y.}_\mathrm{Non-stationary~non-homogeneous~term}
\label{eq:rd}
\end{equation}
Other such identities relating to skin friction are the doubly
integrated boundary layer equation by \citet{xia_huang_cu_cui_2015},
the triply integrated mean vorticity equation by
\citet{yoon_ahn_hwang_sung_2016}, the integrated streamwise momentum
equation with turbulent inertia replacing the Reynolds shear stress by
\citep{zhao_fan_li_2024} and many others
\citep[][etc.]{ricco_skote_2022,elnahhas_johnson_2022}.

Depending on the number of repeated integrations, there are an
infinite number of integral identities relating the skin friction to
different physical process above the wall
\citep{ricco_skote_2022,elnahhas_johnson_2022}.  With each identity,
come correlations between physical processes, each with a different
premultiplied weight. Nevertheless, the mean skin friction is merely
the average over the weighted physical processes, and its value is
independent of the formulation. Attempts to analyse the skin friction
through averaging different physical processes inevitably leads to
ambiguities due to differences in formulations. As an example, the
study of \citet{agostini_leschziner_2019b} used both FIK and RD
identities to analyse the contribution of large-scale motions to the
mean skin friction. Due to the different weights multiplying the
Reynolds shear stress, the results from the two identities are
quantitatively different in terms of percentage contribution to the
mean skin friction by the large-scale motions.

What also remains unclear is whether the upper limit of the
integration should be at infinity in external flows; at the axis or
the plane of symmetry in internal flows; or at an arbitrary wall
normal position. The former two choices, employed by
\citet{vonkarman_1921,fik_2002,xia_huang_cu_cui_2015,renard_deck_2016,
  yoon_ahn_hwang_sung_2016,elnahhas_johnson_2022,ricco_skote_2022},
rely on two potential reasons: (i) it is presumed that skin friction
is influenced by the entire flow above the wall; (ii) it facilitates
the treatment of certain terms that vanish away from the wall.
However, choosing an arbitrary integration limit is also possible and
sometime physically justified. In TBL, \citet{ligrani_moffat_1986}
chose to integrate the boundary layer equation up to a finite height
due to experimentally limited data very far from the wall. As it is
not known {\it{a priori}} which physical processes far away from the
wall contribute to the skin friction, \citet{zhao_fan_li_2024}
investigated how varying the upper limit of the integral terms in
their identity affects contributions to the skin friction in TCF.
Indeed, due to the aforementioned ambiguity in the formulation,
different FIK-like identities may also display distinguishing
behaviours in their variations with the upper limit of the integral,
further obscuring the mean skin friction decomposition.

\subsection{Fluctuating wall shear stress}
The fluctuations of wall shear stress stem from the turbulent flow
above the wall, as they are absent in a laminar steady flow. While
most studies and practices in the turbulent regime concern the mean
skin friction which contributes to the total mean drag, the
fluctuations of wall shear stress also have wide implications, for
example in cardiovascular disease \citep{zhou_etal_2023} and in
structural fatigue modelling of wind turbine blades
\citep{ravikumar_2020}.

The fluctuations of wall shear stress are known to be quite
significant compared to its mean value.  \citet{schlatter_orlu_2010}
compiled fluctuating wall shear stress data from TCF, TPF and TBL and
found that the root-mean-square (rms) of fluctuating wall shear stress
normalised by the mean wall shear stress increases slowly with the
logarithm of friction Reynolds number. At the highest Reynolds numbers
available, the rms of wall shear stress can be more than 40\% of the
mean wall shear stress.  Whether the rms of wall shear stress compared
to its mean value is bounded or not remains an open fundamental
question in wall-bounded turbulent flows.  Assuming that the skin
friction coefficient vanishes at infinite Reynolds number, the
unboundedness of rms of wall shear stress compared to the mean wall
shear stress would imply that for higher Reynolds number applications,
such as large-scale wind turbines, one would have to consider the
fluctuations of local stresses ever more carefully.

Additionally, the increase in the ratio of fluctuating wall shear
stress to the mean wall shear stress with increasing Reynolds number
implies that universal scaling of second-order statistics does not
hold close to the wall. This fact is also evident in the Reynolds
number dependence of the near-wall peak of streamwise momentum
variance.

The fluctuations of wall shear stress are not only characterised by
their intense rms, but also by the nature of their higher order
statistics. Direct numerical simulations (DNS) of turbulent boundary
layers have shown that the wall shear stress follows an approximate
log-normal distribution \citep{diazdaniel_laizet_vassilicos_2017}.
This distribution features a heavy tail that favours the occurrence of
large positive values of instantaneous local wall shear stress,
becoming increasingly extreme with increasing Reynolds number. On the
other side of the distribution, negative wall shear stresses or back
flow events are also observed. This signifies the reversal of local
streamwise velocity inside the viscous sublayer. These extreme
backflow events have been shown to be associated with strong
spanwise motions in the viscous sublayer
\citep{diazdaniel_laizet_vassilicos_2017}.

Experimental measurement of instantaneous and local wall shear stress
is a challenging task that requires forefront experimental techniques
\citep{orlu_vinuesa_2020}. The behaviour of traditional sensors, such
as hot-wires and hot-films, are significantly affected by the solid
surface nearby, which introduces large thermal inertia and drifts the
signals measured \citep{alfredsson_etal_1988}.  Moreover, it is
impossible for these devices to keep track of backflow events due to
their lack of response to flow direction. Only recently have advanced
techniques, such as particle image velocimetry (PIV) and particle
tracking velocimetry (PTV) within the viscous sublayer, enabled
reliable identification of backflow events
\citep{willert_etal_2018,klinner_willert_2025} in good agreement with
DNS data.

To the authors' knowledge, there has been no attempt to date to
describe the fluctuating wall shear stress using the Navier-Stokes
equations to obtain an integral relation a bit like a FIK identity.
\citet{lee_hwang_2025} analysed the fluctuating wall shear stress as
turbulent dissipation at the wall by examining the streamwise
turbulent kinetic energy budget close to the wall.  They found that
turbulent dissipation and transport spectra display universal
behaviour for streamwise wavelength smaller than about 1000 viscous
unit, and contribution for wavelength larger than 1000 viscous unit
and smaller than half height/pipe radius increases with Reynolds
number.

\subsection{Goal of the present work}
A natural attempt to study the wall shear stress and its fluctuations
is by directly employing the streamwise momentum equation (streamwise
component of the Navier-Stokes equations) to describe the wall shear
stress. In the present work, we address the mean skin friction from an
analysis of the plane-average streamwise momentum equation in TCF.
From the the plane-average momentum balance, we derive a simple
identity relating the square mean wall shear stress to the physical
processes above the wall naturally imposed by the Navier-Stokes
equations.
An identity for the variance of the wall shear stress fluctuations is
also derived along the same lines using the plane-fluctuating momentum
balance. The physical processes above the wall which matter in this
identity are also naturally imposed by the Navier-Stokes equations.
The result shows that the fluctuating wall shear stress depends
heavily on the fluctuating acceleration as well as the fluctuating
pressure gradient, both becoming more intense with increasing Reynolds
number. We study the extent from the wall that is required for these
Navier-Stokes-derived processes to determine the mean wall shear
stress and the variance of its fluctuations. We also identify and
study the dominant processes and attempt to explain some of their
features physically.

%%%%%%%%%%%%%%%%%%%%%%%%%%%%%%%%%%%%%%%%%%%%%%%%%%%%%%%%%%%%%%%%%%%%%%
\section{DNS Data set} \label{sec:numerics}
We use the data sets from direct numerical simulation (DNS) of
incompressible FD TCF with constant mass flux at five different
Reynolds numbers: $\Rey_\tau = 235,359,498,953,2012$, where
$\Rey_\tau$ is the friction Reynolds number. The bulk velocity is kept
constant by varying the instantaneous pressure gradient so that it
remains equal to the sum of surface frictions on the two walls. The
data sets are obtained with DNS performed using the Navier-Stokes
solver \texttt{diablo} \citep{bewley_2014}, employing the
Fourier-Galerkin method with a 2/3 de-aliasing rule in the
wall-parallel directions $x$ and $z$, and a second-order finite
difference method in the wall-normal direction $y$. The temporal
discretisation of the solver is based on the fractional-step algorithm
by \cite{kim_moin_1985}, with implicit treatment of viscous terms
using the Crank–Nicolson scheme and explicit treatment of the
remaining terms using a low-storage third-order Runge–Kutta scheme.
It is the same data set used by \citet{yin_hwang_vassilicos_2024},
except that the R2000 simulation has been continued up to
$T_\mathrm{int} u_\tau/\delta = 145.79$ for better convergence of
statistics (see exact definition of $u_\tau$ in sub-section 3.1). We
denote the streamwise, wall-normal, and spanwise directions by
$(x,y,z)$, with the corresponding velocity field components $(u,v,w)
\equiv (u_1, u_2, u_3)$.  The size of the simulation domain is set to
be $L_x = \pi \delta$, $L_y = 2\delta$, and $L_z = \pi \delta/2$, where $\delta$ is the half channel height. The spanwise extent of the
simulation box is
comparable to the minimal unit for the self-sustaining process of the
largest energy-containing structures as per \cite{hwang_cossu_2010c}
and \cite{hwang_bengana_2016}.  Adhering to previous studies, the size
of the computational domain is deliberately chosen to be small in
order to capture single entities of the large-scale energy-containing
structures without significantly modifying the mean statistics
\citep{lozano-duran_jimenez_2014a}.  Details of the DNS data set and
its naming convention can be found in table \ref{tab:setup}.

\begin{table}
	\begin{center}
	\begin{tabular}{c c c c c c c c}
		Name & $\Rey_\tau$	& $N_x \times N_y \times N_z$	& $\Delta x^+$	&	$\Delta y^+$ & $\Delta z^+$	& $T_\mathrm{int} u_\tau/\delta$ & $N_t$\\ [3pt]
		
		R230 & 235	& $96 \times 193 \times 96$		& 11.52			& 0.53 -- 4.53  & 5.76			& 183.20		& 3000\\
		R360 & 359	& $144\times 257 \times 144$		& 11.74			& 0.61 -- 5.19	& 5.87			& 172.24		& 3000\\
		R500 & 498	& $192\times 289 \times 192$		& 12.22			& 0.75 -- 6.41	& 6.11			& 164.46		& 3000\\
		R950 & 953	& $384\times 513 \times 384$		& 11.69			& 0.80 -- 6.91	& 5.85			& 151.55		& 3000\\
		R2000& 2012	& $768\times 1025\times 768$		& 12.31			& 0.83 -- 7.29	& 6.16			& 145.79		& 3151
	\end{tabular}
	  \caption{Numerical setup of DNS in FD TCF. $N_x, N_y$ and $N_z$ indicate the numbers of grid points in the streamwise, wall-normal and spanwise directions respectively; $\Delta x^+, \Delta y^+$ and $\Delta z^+$ indicate grid spacings, after de-aliasing, in inner units; $T_\mathrm{int}$ denotes the integration time after the initial transient and $N_t$ denotes the number of instantaneous snapshots stored for post-processing.}
	  \label{tab:setup}
	 \end{center}
\end{table}

%%%%%%%%%%%%%%%%%%%%%%%%%%%%%%%%%%%%%%%%%%%%%%%%%%%%%%%%%%%%%%%%%%%%%%
\section{Plane-average wall shear stress} \label{sec:mean}
As pointed out in \S\ref{sec:intro}, we aim to study the wall shear
stress using the Navier-Stokes equations.  The wall shear stress is a
consequence of fluid motion above the wall, the adherence of fluid at
the wall, and the effect of viscous momentum diffusion in the vicinity
of the wall.  It is therefore natural to extract the wall shear stress
directly from the diffusive term of the Navier-Stokes equations. The
only way that the Navier-Stokes equations can be recast so as to give
formulae for fluctuating wall shear stress moments in terms of fluid
flow quantities above the wall is by multiplying the fluctuating
momentum equation by a power of the kinematic viscosity times the
fluctuating streamwise velocity gradient in the wall normal direction,
and then integrating in the wall-normal direction from the wall to
some distance from the wall. Depending on the order of the term
multiplying the Navier-Stokes equations (the value of the
aforementioned power), one obtains an equation that exactly describes
any order statistics of the fluctuating wall shear stress as explained
in Appendix \ref{sec:A_nth}. In this work, we apply this procedure to
the variance of the fluctuating wall shear stress but start by
studying the mean wall shear stress. We apply a similar procedure to
obtain a closed form equation for the square of the mean wall
shear-stress as explained in the following subsection. The present
study is therefore limited to the mean wall shear stress and the
variance of wall shear stress, i.e. the first and second order
statistics.

\subsection{Derivation of plane-average mean equation}
\begin{figure}
\centering
\includegraphics[scale=1]{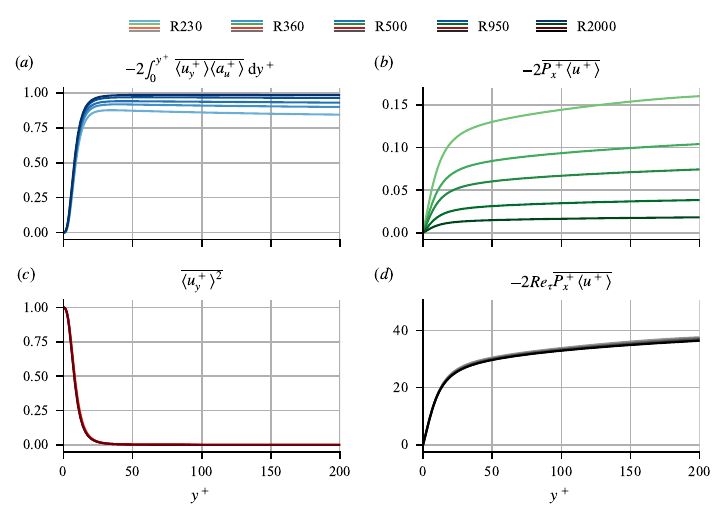}
\caption{Contribution to the square plane-average wall shear stress in equation (\ref{eq:pavg}).}
\label{fig:pavg}
\end{figure}

We consider the plane-averaged streamwise Navier-Stokes equation:
\begin{equation}
\langle a_u \rangle = -P_x+\nu\langle u_{yy}\rangle,
\label{eq:avgns}
\end{equation}
where $a_u$ is the fluid particle acceleration comprised of Eulerian
acceleration plus non-linear advection, $P_x$ is the $(x,z)$
plane-averaged pressure gradient (implicitely normalised by mass
density), $u_{yy}$ stands for $\frac{\partial^2}{\partial y^2} u $,
$\nu$ is the kinematic viscosity, and the brackets represent averaging
in a wall-parallel plane, i.e. $\langle A \rangle = \frac{1}{L_x L_z}
\iint A ~\mathrm{d}x\mathrm{d}z$. (The wall-parallel diffusion term
averages to zero due to periodicity in the wall-parallel boundary
conditions.)  We now multiply both sides of equation (\ref{eq:avgns})
at arbitrary $y$ by $\nu\langle u_y \rangle(y) =
\nu\frac{\partial}{\partial y}\langle u \rangle(y)$ at a given $y$,
and integrate the equation in the $y$ direction from $y=0$ at the wall
to some arbitrary $y$:
\begin{eqnarray}
\nu\int_0^y \langle u_y \rangle\langle a_u \rangle ~\mathrm{d}y' &=& -\nu\int_0^y \langle u_y \rangle P_x ~\mathrm{d}y' + \frac{1}{2}\left[\nu^2\langle u_y \rangle^2(y,t) - \nu^2\langle u_y \rangle^2(y=0) \right]\nonumber \\
\Rightarrow	\langle\tau_w\rangle^2(t) &=& -2\nu\int_0^y \langle u_y \rangle\langle a_u \rangle ~\mathrm{d}y' -2\nu P_x \langle u \rangle(y,t) + \nu^2\langle u_y \rangle^2(y,t).
\label{eq:pavg}
\end{eqnarray}
We obtain this exact formula (\ref{eq:pavg}) for the time-fluctuating
plane-averaged wall shear stress as a sum of three different
quantities that depend on $y$ distance from the wall. {\color{black} As the left-hand-side of the equation (\ref{eq:pavg}) is independent of $y$, t}he
$y$-dependencies of the right hand side terms cancel at every $y$.
When taking $y=0$ on the right hand side, this relation reduces to the
definition of the local wall shear stress, i.e. $\tau_w (x,z,t) \equiv
\nu u_y (x, y=0,z,t)$, with vanishing shear-acceleration term and
pressure-velocity term.

Defining the time average $ \overline{A} = \lim_{T\to\infty}
~\frac{1}{T}\int_0^T ~ A (t) ~\mathrm{d}t $ for any variable $A(t)$,
the time-averaged terms on the right hand side of equation
(\ref{eq:pavg}) are plotted in figure \ref{fig:pavg}($a,b,c$).  When
presented with superscript $(\cdot)^+$, the results are normalised by
the usual inner units, i.e. the friction velocity $u_\tau$ defined
from the space-time average wall shear stress $u_\tau^2 =
\overline{\langle \tau_w \rangle}$, and the viscous wall unit
$\delta_\nu = \nu / u_\tau$.
From (\ref{eq:pavg}), the sum of the three quantities plotted in
figure \ref{fig:pavg}($a,b,c$) equals
$\overline{\langle\tau_w\rangle^2}/u_{\tau}^{2}$ which is very close
to 1 (see following paragraph).

Due to the minimal computational domain, the plane-average wall shear
stress $\langle \tau_w \rangle(t)$ fluctuates in time.  Its time
fluctuation in terms of rms is less than 5\% of its time average value
for all five Reynolds numbers.  The time-average squared wall shear stress
therefore contains less than 0.25\% deviation, which is
insignificant compared to the square of the time-average wall shear stress.
Numerical integration in the wall-normal direction is performed with
cubic spline, and the resulting numerical error is within 1\% of the
square mean WSS.

\subsection{Estimate of wall shear stress and Reynolds number dependencies}\label{sec:3.2}

\begin{figure}
\centering
\includegraphics[scale=1]{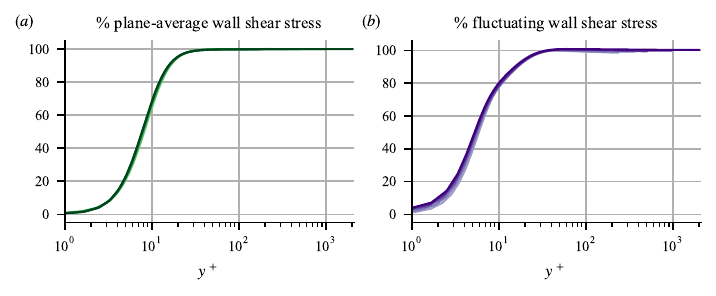}
\caption{Percentage contribution from shear-acceleration plus
  shear-pressure-gradient to ($a$) plane-average and ($b$) fluctuating
  wall shear stresses. Results from five Reynolds numbers collapse
  rather well.}
\label{fig:percent}
\end{figure}

The time-average square mean shear stress $\overline{ \langle {u_y^+}
  \rangle^2}$ normalised by inner units
shown in figure \ref{fig:pavg}($c$) is near-equal to 1 at the wall
and reduces monotonically as $y$ increases further away from the wall.
At $y^+\approx 54$, $\overline{ \langle {u_y^+} \rangle^2}$ drops
below 0.2\% of time-average square wall shear stress $\overline{
  \langle \tau_w\rangle^2}/u_{\tau}^{4}$ for all five Reynolds
numbers.  Similar to the $\delta_{99}$ definition for the turbulent
boundary layer thickness, we may define a position $\delta_w =
\delta_{0.2\%} \equiv 54 \delta_\nu$ as a cut-off wall normal
position. Changing the cut-off threshold, the 1\% cut-off position is
around $31 \delta_\nu$ for example. This cut-off position identifies
the $y$-location above which only the shear-acceleration term plotted
in figure \ref{fig:pavg}($a$) and the pressure-velocity term plotted
in figure \ref{fig:pavg}($b$) effectively (if not overwhelmingly)
contribute to $\overline{ \langle \tau_w\rangle^2}$.
{\color{black}(Note that the local contribution from the molecular shear stress
 vanishes beyond this cut-off position, as plotted in 
figure \ref{fig:pavg}($c$).)}
{\color{black} The dominance of these two contributions to $\overline{ \langle \tau_w\rangle^2}$} is confirmed
by figure \ref{fig:percent}($a$) where we plot the percentage
contribution to $\overline{ \langle \tau_w\rangle^2}$ from these two
terms as a function of the upper $y$ limit of the integral. It is
found that the corresponding curves for the five Reynolds numbers
collapse rather well, but not perfectly. This result suggests that the
buffer layer encodes within it all the necessary information required
to calculate the plane-average WSS.

{\color{black}To better appreciate this result, it may help to spell out the
differences between equation (\ref{eq:pavg}) and the usual integrated
mean momentum equation obtained by directly integrating
(\ref{eq:avgns}) over $y$:
\begin{equation}
\langle \tau_w\rangle = -\int_{0}^{y} \langle a_u \rangle ~\mathrm{d}y'
- P_x y +\nu\langle u_{y}\rangle (y,t).
\label{eq:intavgns}
\end{equation}
The integrand $\langle u_y \rangle\langle a_u \rangle$ in the first
term of the right hand side of (\ref{eq:pavg}) decays to zero with
increasing $y'$ whereas the integrand $\langle a_u \rangle$ in the
first term of the right hand side of (\ref{eq:intavgns}) does not. In
particular, time averaging of both sides of (\ref{eq:intavgns}) leads
to $\overline{\langle \tau_w\rangle} = -\overline{\langle u v\rangle}
(y) - \overline{P_x} y +\nu\overline{\langle u_{y}\rangle} (y)$ which
makes it clear that $\int_{0}^{y} \overline{\langle a_u \rangle}
\mathrm{d}y' = \overline{\langle u v\rangle} (y)$ does not tend to
become independent of $y$ with increasing $y$ as it would if
$\overline{\langle a_u \rangle}$ decayed to $0$ fast enough with
increasing $y'$. Hence whilst the first term on the right hand side of
(\ref{eq:pavg}) does not grow with $y$ above $\delta_w$, the first
term on the right hand side of (\ref{eq:intavgns}) does, and this
is particularly clear from the time averaged form of equation
(\ref{eq:intavgns}) where $-\overline{\langle u v\rangle} (y)$
continues growing till $y$ reaches $y_{m} \sim \sqrt{\delta_{\nu}
  \delta}$. Indeed, the wall distance $y_{m} \sim \sqrt{\delta_{\nu} \delta}$
where $-\overline{\langle u v\rangle}$ is maximal follows directly by
assuming a log law for $\overline{\langle u \rangle}$ \citep[see e.g.][]{afzal_1984,panton_2007,lee_moser_2015,hwang_lee_2020}. This length
scale $\sqrt{\delta_{\nu} \delta}$ characterises the mesolayer \citep{afzal_1984} and is larger than $\delta_{w}$ which scales with
$\delta_{\nu}$. Note that the linear dependence on $y$ of the pressure
term on the right hand side of (\ref{eq:intavgns}) is essential for
obtaining this mesolayer length scale and that the dependence on $y$
of the pressure term on the right hand side of (\ref{eq:pavg}) is much
weaker than linear. Given these important differences between the
integrated momentum equation (\ref{eq:intavgns}) and our equation
(\ref{eq:pavg}), it is not possible to use the integrated mean
momentum equation (\ref{eq:intavgns}) to encode within the buffer
layer all necessary information required to calculate the
plane-average WSS as it is with our equation (\ref{eq:pavg}).}

We now focus on the terms on the right hand side of equation
(\ref{eq:pavg}). While $\overline{ \langle {u_y^+} \rangle^2}$ shows
little Reynolds number dependence close to the wall (figure
\ref{fig:pavg}$c$), both the shear-acceleration and the
pressure-velocity terms show significant Reynolds number effects
(figure \ref{fig:pavg}$a,b$). The latter's Reynolds number dependence
can be estimated as follows.
In a constant mass flux channel, the plane-average pressure gradient
balances the sum of the plane integrals of wall shear stresses on the
two walls instantaneously. By assuming that the plane-average pressure
gradient does not correlate heavily in time with the plane-average
velocity profile, one can estimate the mean pressure gradient as the
squared friction velocity divided by the channel half height, i.e.
\begin{equation}
-\nu  \overline{P_x \langle u \rangle} /u_\tau^4 \approx ~ 
\frac{\nu \overline{\langle u \rangle} u_\tau^2 }{  u_\tau^4 \delta}
=~\frac{\overline{\langle u \rangle}}{u_\tau} \frac{\nu}{\delta u_\tau}
=~\frac{U^+(y^{+})}{Re_\tau},
\end{equation}
where $U^+(y^{+})$ is the mean velocity profile scaled in inner units
which may be assumed to be invariant with Reynolds number in the
near-wall region.  We therefore plot in figure \ref{fig:pavg}($d$) the
pressure velocity term multiplied by $Re_\tau$. The product collapses
approximately for all five Reynolds numbers. The slight mismatch
between the proposed scaling and the data from the current simulation
may be due to two main possible causes: time correlations between the
time fluctuations of the plane-average pressure gradient and the
plane-average velocity; and residual weak Reynolds number effects on
the mean velocity profile. We note that the above balance would be
exact in a constant pressure gradient channel (with $\approx$ replaced
by $=$).

At high enough Reynolds number, we therefore have the following
estimation of the time-average square mean wall shear stress in terms
of the product between the mean shear and the plane-average
acceleration integrated from the wall to some finite distance from the
wall, plus a correcting pressure-velocity term that is proportional to
$1/\Rey_\tau$ (this estimate is acceptable even without
time-averaging):
\begin{equation}
\overline{\langle \tau_w\rangle ^2} \approx	-2\nu\int_0^{\delta_w} \overline{\langle u_y \rangle \langle a_u \rangle} ~\mathrm{d}y' + O(1/\Rey_\tau).
\label{eq:pavgapprox}
\end{equation}

This result signifies the existence of a cut-off wall-normal distance
$y = \delta_w$ which seems to scale with $\delta_\nu$ and beyond which
there is no more additive contribution from the shear-acceleration
term to $\overline{\langle \tau_w\rangle ^2}$. The pressure gradient
contribution vanishes at a $1/\Rey_\tau$ rate at large Reynolds
numbers. We note that such interpretation is limited to the derived
identity, and does not necessarily apply to other identities, for
reasons explained in \S\ref{sec:fik}.  Once sufficient information
regarding the plane-average acceleration and shear near the wall
(viscous and buffer layers) are gathered, it is possible to
estimate $\overline{\langle \tau_w\rangle ^2}$ exclusively based on
near-wall data. Note, however, that $\delta_w /\delta_\nu$ may have a
weak dependence on $\Rey_\tau$ which cannot be captured with the range
of Reynolds numbers of our DNS data sets and which may imply that
information regarding the plane-average acceleration and shear may
also be required from above the buffer layer at some enormous value of
$\Rey_\tau$ .

\begin{figure}
\centering
\includegraphics[scale=1]{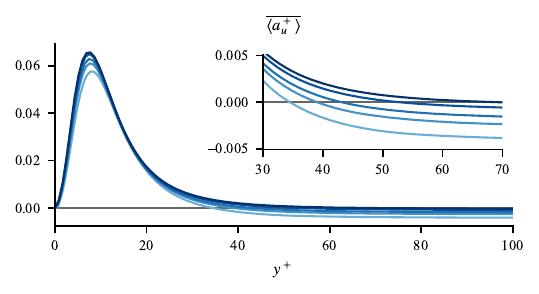}
\caption{Mean acceleration profile.}
\label{fig:au}
\end{figure}

As all indications to date are that the mean skin friction coefficient
has a weak dependence on $\Rey_\tau$ with increasing $\Rey_\tau$, one
may also expect $\overline{\langle \tau_w\rangle ^2}$ to have such an
asymptotic dependence. This dependence may come either from a
$\Rey_\tau$ dependence of $\delta_w /\delta_\nu$ which is beyond our
reach in this paper, or from the $\Rey_\tau$ dependence of the
integrand in the integral on the right hand side of
(\ref{eq:pavgapprox}), or both. The plane-average shear profile $
\langle u_y \rangle$ may be considered as almost fully scaling in inner 
units near the wall. {\color{black} (This is reminiscent of the law of the wall, which states that the time- and plane-average shear profile $\overline{\langle u_y \rangle}$ fully scales in inner units near the wall.)}
 However, the plane-average acceleration $\langle a_u
\rangle$ (figure \ref{fig:au}) is composed of the plane-average
Eulerian acceleration and the Reynolds shear stress gradient, the
latter clearly depending on Reynolds number. Even in the minimal
computational domain, the plane-average Eulerian acceleration is less
than $1\%$ of the plane-average Reynolds shear stress gradient.  The
gradient of Reynolds shear stress is known to cross zero (local
maximum of Reynolds shear stress) at a $y^+$ that increases with
Reynolds number. Reynolds number dependencies in magnitude and in
zero-crossing location are clearly shown in figure \ref{fig:au} where
we plot the wall-normal profile of $\overline{\langle a_u \rangle}$ in
inner coordinates. If $\delta_w /\delta_\nu$ is independent of
$\Rey_\tau$ then all the information required to calculate
$\overline{\langle \tau_w\rangle ^2}$ from the fluid flow can be found
between the wall and the outer edge of the buffer layer for any
Reynolds number. However, this thin near-wall region will then need to
be Reynolds number dependent if $\overline{\langle \tau_w\rangle ^2}$
is Reynolds number dependent, and this Reynolds number dependence will
reflect a dependence of this thin near-wall region on the outer flow.

%%%%%%%%%%%%%%%%%%%%%%%%%%%%%%%%%%%%%%%%%%%%%%%%%%%%%%%%%%%%%%%%%%%%%%
\begin{figure}
\centering
\includegraphics[scale=1]{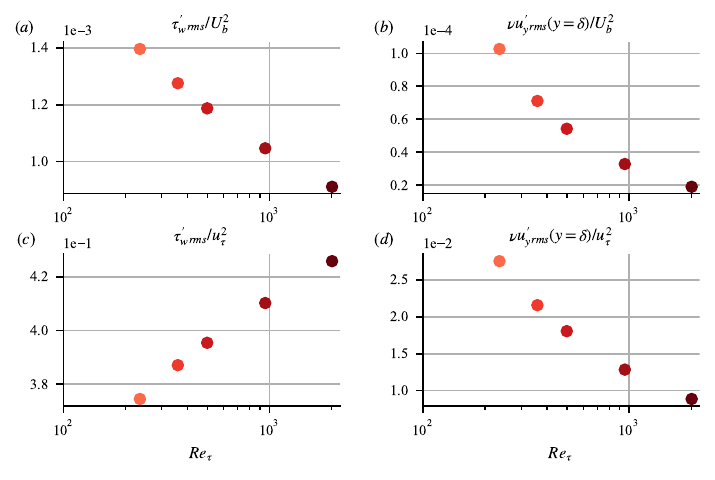}
\caption{rms of fluctuating wall shear stress and local shear at channel centre normalised by bulk quantities ($a,b$) and by inner quantities ($c,d$)}
\label{fig:uyp}
\end{figure}
\section{Wall shear stress fluctuations}

Our method for analysing local quantities at the wall relies on the
integration of the wall-normal diffusion term appearing in the
Navier-Stokes equations, whether it is plane-averaged or
plane-fluctuating. Our analysis of the fluctuating wall shear stress
directly from the Navier-Stokes equations leads to a relation in terms
of covariances that is neither an energy nor a momentum balance.
Without any assumptions, this relation includes the covariances that
are most direct and natural from the Navier-Stokes equations for the
mean and fluctuations of square wall shear stress.

A lesson learnt from the analysis of cut-off height in the previous
section is that nearly all the information needed to determine the
mean wall shear stress can be found in the viscous and buffer layers
at the Reynolds numbers of the present DNS data. The question
naturally arises whether such cut-off wall-normal height also exists
for the fluctuating wall shear stress.

To derive a formula for the variance of fluctuating wall shear stress
from the Navier-Stokes equations, we first multiply the fluctuating
wall-normal diffusion term in that equation with the fluctuating shear
stress at the same $y$, average in $x$-$z$ plane, and then integrate
in the $y$ direction from the wall to an arbitrary $y$:
\begin{equation}
\nu^2\int_0^{y} \langle u_y' u_{yy}' \rangle ~\mathrm{d}y' = \frac{\nu^2}{2}\int_0^{y} \frac{\partial}{\partial y} \langle u_y' u_{y}' \rangle ~\mathrm{d}y' =  \frac{\nu^2\langle {u_y'}^2\rangle(y,t)}{2} - \frac{\langle {\tau_w'}^2 \rangle(t)}{2},
\label{eq:uyp}
\end{equation}
where the prime symbol denotes the plane-fluctuating part of a
quantity, i.e.: $A'(x,y,z,t) = A(x,y,z,t) - \langle A
\rangle(y,t)$. We see how this operation leads to the appearance of
the variance of fluctuating wall shear stress $\langle {\tau_w'}^2
\rangle(t)$ so that when applied to all the terms of the Navier-Stokes
equations at the start of sub-section 4.1, this operation yields a
formula for $\langle {\tau_w'}^2 \rangle(t)$ in terms of fluid flow
statistics above the wall. However, an additional interesting feature
of equation (\ref{eq:uyp}) is the appearance of the variance of the
local shear stress which is not zero at the channel centre, in
contrast to the mean shear which is zero at the channel centre due to
symmetry. Before proceeding with the Navier-Sokes equations in the
following sub-section, we compare the Reynolds number dependencies of
the two variances on the right hand side of equation (\ref{eq:uyp})
for $y=\delta$. 

If normalised by $\delta/U_b$, the time-average rms of the fluctuating
shear ${u'_y}_{rms}(y) \equiv \overline{\sqrt{\langle
    {u_y'}^2\rangle}}(y)$ increases with increasing $\Rey_\tau$ both
at $y=0$ and $y=\delta$. However these increases are slower than
$1/\nu$ because ${\tau'_w}_{rms} \equiv \nu {u'_y}_{rms}(y=0)$ and
$\nu {u'_y}_{rms}(y=\delta)$ both decrease with increasing $\Rey_\tau$
when normalised by $U_{b}^{2}$ as shown 
in figure \ref{fig:uyp}(a,b) for the five Reynolds numbers available.
It is noticeable that $\nu {u'_y}_{rms}(y=\delta)/U_{b}^{2}$ decreases
much faster with increasing $\Rey_\tau$ than
${\tau'_w}_{rms}/U_{b}^{2}$.  This difference appears more
dramatically when normalising by $u_{\tau}^{2}$. Figure
\ref{fig:uyp}($c$) shows that ${\tau'_w}_{rms}/u_{\tau}^{2}$ increases
with increasing $\Rey_\tau$, confirming the observation in
\cite{schlatter_orlu_2010}, whereas figure \ref{fig:uyp}($d$) shows
that $\nu {u'_y}_{rms}(y=\delta)/u_{\tau}^{2}$ decreases with
increasing $\Rey_\tau$. 
Similarly to our treatment of the mean wall shear stress, this
observation could contribute to the existence of a cut-off height for
the fluctuating wall shear stress equation, an issue which we address
in the following sub-section.

\subsection{Formula for the wall shear stress variance and time-average statistics}
We consider the plane-fluctuating streamwise momentum equation:
\begin{equation}
a_u' = -p_x'+\nu \nabla^2 u'.
\label{eq:flucns}
\end{equation}
We first multiply the above equation by $\nu u_y'$, average over the
$x$-$z$ plane, integrate in the wall-normal direction from the wall to
some arbitrary $y$ and using equation (\ref{eq:uyp}) we obtain
\begin{equation}
\langle\tau_w'^2\rangle = 
-2\nu  \int^y_0 \langle  u'_y a_u'\rangle ~\mathrm{d}y'
-2\nu  \int^y_0 \langle  u'_y p_x' \rangle ~\mathrm{d}y'
+ 2\nu^2\int^y_0 \langle u'_y u_{//}' \rangle~\mathrm{d}y'
+ \nu^2 \langle {u'_y} ^2\rangle, \label{eq:pfluc}
\end{equation}
where $p_x'$ is the fluctuating streamwise pressure gradient
(implicitely normalised by the mass density), $u_{//}' =
u_{xx}'+u_{zz}'$ is the sum of the second derivatives in $x$ and $z$
directions of the streamwise velocity, and $\tau_w' \equiv \nu
u_y'(y=0)$ is the fluctuating wall shear stress that appears from the
vertical momentum diffusion term.  To the authors' knowledge, this is
the first exact relation describing the spatially fluctuating wall
shear stress directly using the Navier-Stokes equations.

\begin{figure}
\centering
\includegraphics[scale=1]{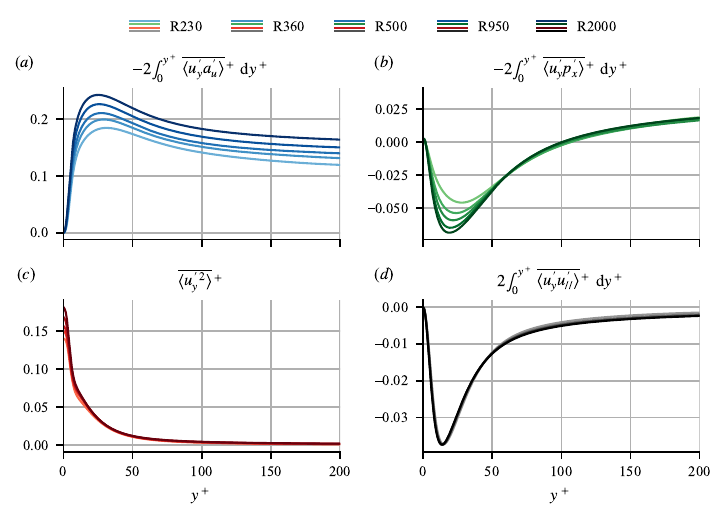}
\caption{Contributions to the square plane-fluctuating wall shear stress in equation (\ref{eq:pfluc}).}
\label{fig:fluc}
\end{figure}

The wall-normal profile of each time averaged budget term on the right
hand side of equation (\ref{eq:pfluc}) is plotted in figure
\ref{fig:fluc} normalised in inner units.
The major contribution to the wall shear stress fluctuations far
enough from the wall is the positive contribution from the integrated
shear-acceleration covariance attenuated by the negative contribution
from the integrated shear-pressure-gradient covariance at normalised
wall distances $y^+$ below 100. The latter contribution turns from
negative to positive around $y^+ \approx 100$ but remains relatively
minor at $y^+ > 100$. The two terms on the right hand side of equation
(\ref{eq:pfluc}) stemming from viscous diffusion diminish in magnitude
with wall distance and appear to quickly become negligible as shown in
figure \ref{fig:fluc}($c,d$). Whilst the profiles of these two viscous
diffusion-related terms collapse rather well in inner units (except
very near the wall for $\nu^2 \langle {u'_y} ^2\rangle$) the two main
contributions to $\langle\tau_w'^2\rangle$ display a significant
increase in magnitude with increasing $\Rey_{\tau}$ in figure
\ref{fig:fluc}($a,b$) where they are plotted in inner units. The
magnitude of the negative integrated shear-pressure-gradient
covariance increases around $y^+\approx 20$ with increasing Reynolds
number, as opposed to the diminishing contribution of the
plane-average pressure-velocity term to the mean wall shear stress.

The major contribution to both the plane-average and the fluctuating
wall shear stress comes from an integral involving shear and
acceleration, with some correction from the term involving pressure.
We plot in figure \ref{fig:percent}($b$) the time-integrated sum of
the $y$-integrated shear-acceleration covariance and the
$y$-integrated shear-pressure-gradient covariance as a percentage
contribution to $\overline{\langle\tau_w'^2\rangle}$. The percentage
is given as a function of the upper $y$ limit of the integrals. Even
though there is no plateau reaching behaviour in figure
\ref{fig:fluc}($a,b$) as there is in figure \ref{fig:pavg}($a$) so that
both the shear-acceleration and the shear-pressure-gradient terms
contribute significantly to the fluctuating wall shear even beyond
$y^+ > 200$, the percentage plotted in figure \ref{fig:percent}($b$)
does reach a plateau beyond $y^+\approx 30$. This observation stems
from the approximate balance between fluctuating acceleration and
fluctuating pressure far from the wall where fluctuating viscous
diffusion is negligible (see \S\S \ref{sec:p+au}). As a result, we may
expect the following estimate for the variance of the fluctuating wall
shear stress:
\begin{equation}
\overline{\langle {\tau_w'}^2\rangle } \approx	-2\nu\int_0^{\delta_{w'}} \overline{\langle u_y' a_u' \rangle} ~\mathrm{d}y'-2\nu\int_0^{\delta_{w'}} \overline{\langle u_y' p_x' \rangle} ~\mathrm{d}y'.
\label{eq:flucapprox}
\end{equation}

This estimate parallels the estimate (\ref{eq:pavgapprox}) for the
mean wall shear stress. For the Reynolds numbers accessible by our FD
TCF data, the Navier-Stokes equations implies that one can predict the
wall shear stress mean and variance from statistics involving
mean/fluctuating shear, acceleration and pressure gradient between the
wall and the upper edge of the buffer region. Similar to the estimate
(\ref{eq:pavgapprox}), the choice of the upper limit $\delta_{w'}$ of
the integral in $y$ depends on the threshold we chose for deciding
that the percentage in figure \ref{fig:percent}($b$) is close enough
to 100\%. Comparing figures \ref{fig:percent}($a$) and
\ref{fig:percent}($b$), the percentage contribution to the fluctuating
wall shear stress shows slightly larger variability with varying
Reynolds numbers. This observation suggests that $\delta_{w'}$ in
equation (\ref{eq:flucapprox}) may have a slightly stronger dependence
on $Re_\tau$ than $\delta_w$ in equation
(\ref{eq:pavgapprox}). However, the Reynolds number dependence of
$\overline{\langle {\tau_w'}^2\rangle }$ shown in figure
\ref{fig:uyp}($a,c$) may also arise from the integrands inside the
integrals in equation (\ref{eq:flucapprox}). Even so, equation
(\ref{eq:flucapprox}) suggests that the variance of the fluctuating
WSS is fully encoded within the buffer layer, i.e. that it is possible
to obtain it from information within it.

\begin{figure}
\centering
\includegraphics[scale=1]{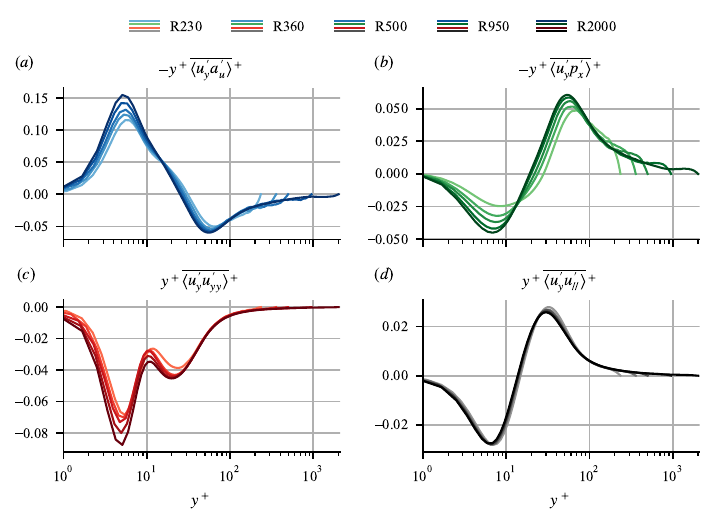}
\caption{Premultiplied integrand contribution to square fluctuating plane-average wall shear stress.}
\label{fig:integrand}
\end{figure}
In figure \ref{fig:integrand}($a$) and ($b$) we plot these two
integrands in inner units as functions of $y^+$. As the vertical axis
is linear and the horizontal axis is logarithmic, we premultiply them
by $y^+$ to ensure that the integral over $\log y^{+}$ is the same.
Plotted in the same way in figure \ref{fig:integrand}($c$) is the
spatial covariance between local shear and wall-parallel diffusion of
streamwise fluctuating velocity, whose integral gives the difference
between fluctuating wall shear stress and local-in-$y$ streamwise shear
(see relation \ref{eq:uyp}). The sum of the four covariances in figure
\ref{fig:integrand}($a,b,c,d$) is zero at each $y^+$, as they are
exactly balanced by the streamwise momentum equation.
It is not until the channel centre $y^+ = \Rey_\tau$, the plane of
symmetry, that the differential contribution from the
shear-acceleration covariance becomes zero.  The premultiplied
shear-acceleration covariance (figure \ref{fig:fluc}$a$) and
shear-pressure-gradient covariance (figure \ref{fig:fluc}$b$) beyond
$y^+ = 30$ seem to cancel each other far enough from the wall, sharing
similar positive/negative peak location around $y^+ \in [50,60]$ as
well as peak magnitudes.  This observation is consistent with the
integral estimate (\ref{eq:flucapprox}) and the integral profiles
plotted in figure \ref{fig:fluc}. Concerning Reynolds number
dependence, a discussion that parallels the discussion at the end of
\S\S \ref{sec:3.2} could be repeated here for $\overline{\langle
  {\tau_w'}^2\rangle }$ and refering to figure \ref{fig:fluc}($a,b$)
rather than $\overline{\langle \tau_w\rangle ^2}$ and refering to
figure \ref{fig:au}. 

\subsection{Pressure and acceleration}\label{sec:p+au}

\begin{figure}
\centering
\includegraphics[scale=1]{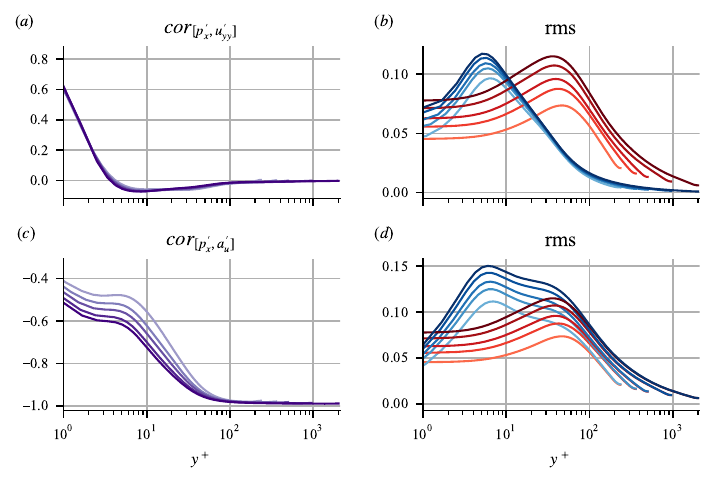}
\caption{Pressure effect. ($a$) and ($c$): correlation of streamwise
  pressure gradient, vertical viscous diffusion and acceleration. In
  ($b$) and ($d$), red: rms of pressure gradient fluctuation. In ($b$)
  blue, rms of vertical viscous diffusion. In ($d$) blue, rms of
  streamwise acceleration. All rms are normalised in inner units.}
\label{fig:pxau}
\end{figure}

The fluctuating acceleration, pressure gradient and viscous diffusion
are related exactly by the fluctuating streamwise momentum equation
(\ref{eq:flucns}). Approaching the wall as $y^+ \to 0$, velocity and
acceleration vanish, resulting in the balance $p_x' \approx \nu
u_{yy}'$ in some typical sense sufficiently close to the wall. Further
away from the wall, i.e where $y^+ \gg 1$, the viscous diffusion as
the solenoidal part of the force (normalised by mass density)
balancing the acceleration, becomes less and less dominant. We expect
therefore $a_u' \approx -p_x'$ to typically hold away from the wall,
thereby explaining the self-cancelling behaviour in figure
\ref{fig:integrand}($a,b$). To examine the validity of these two
approximations, we plot in figure \ref{fig:pxau}($a$) and
\ref{fig:pxau}($c$) the correlation coefficient between pressure
gradient and vertical viscous diffusion of streamwise momentum, and
the correlation coefficient between pressure gradient and streamwise
acceleration respectively.  For zero-mean variables $A(x,y,z,t)$ and
$B(x,y,z,t)$ fluctuating in space and time, we define a spatial
correlation coefficient as:
\begin{equation}
cor_{[A,B]}(y) = \overline{\left(\frac{\langle AB \rangle}{\sqrt{\langle{A}^2\rangle}\sqrt{\langle{B}^2\rangle}}\right)}.
\label{eq:cor}
\end{equation}

Inside the viscous sublayer and above the wall, the correlation
coefficient between pressure gradient and vertical diffusion (figure
\ref{fig:pxau}$a$) reduces quickly with distance from the wall.  The
correlation coefficient between pressure gradient and acceleration
(figure \ref{fig:pxau}$c$) remains negative inside the viscous
sublayer, showing very significant anti-correlation. These two
observations combined indicate that, inside the viscous sublayer
$y^+<10$, both viscous diffusion and pressure gradient contribute to
the acceleration, and the three quantities are comparable to each
other in their rms (figure \ref{fig:pxau}$b,d$) and correlations.  The
respective rms values shown are consistent with the study of
\citet{yeo_kim_lee_2010} (see their figures 3 and 4). Significant
deviations to the pressure-diffusion balance, which is actually exact
at the wall, already occur extremely close to the wall as acceleration
quickly develops.

In the range of $10<y^+<100$, the rms of vertical diffusion quickly
reduces and the correlation between streamwise pressure gradient and
acceleration quickly develops, attaining nearly $-1$ as an almost
perfect anti-correlation.  Note that this almost perfect
anti-correlation can be found in isotropic turbulence even at moderate
Reynolds number (e.g. see figure 4 in
\citet{tsinober_vedula_yeung_2001}). Above $y^+>100$, the rms values
of pressure and acceleration almost coincide. The self-cancellation
between pressure gradient and acceleration is therefore supported by
the almost perfect anti-correlation and coinciding rms values, and
also evidenced by the absence of viscous diffusion far from the wall.

\subsection{Taylor's hypothesis}
\begin{figure}
\centering
\includegraphics[scale=1]{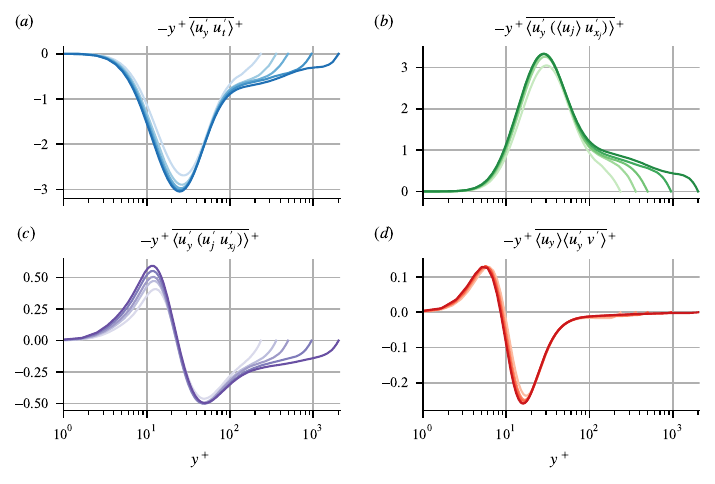}
\caption{Decomposed acceleration contributions.}
\label{fig:advec}
\end{figure}
Among the four premultiplied covariances presented in figure
\ref{fig:integrand}, the shear-acceleration covariance provides the
largest contribution to $\overline{\langle {\tau_w'}^2\rangle }$. We
investigate this primary contribution to the wall shear stress
fluctuation variance by applying the Reynolds decomposition and
decomposing acceleration into five parts:
\begin{equation}
a_u' = u_t' + \langle u_j \rangle u_{x_j}' + u_j' u_{x_j}' + u_j'
\langle u_{x_j} \rangle  -\langle u_j' u_{x_j}' \rangle,
\label{eq:audecomp}
\end{equation}
where $u_t'=\frac{\partial}{\partial t} u'$ is the plane-fluctuating local 
acceleration in the
Eulerian frame of reference, i.e. fixed with respect to the wall;
$\langle u_j \rangle u_{x_j}' = \langle u \rangle u_{x}'+\langle w
\rangle u_{z}'$ ($u_{x}'$ and $u_{z}'$ stand for
$\frac{\partial}{\partial x} u'$ and $\frac{\partial}{\partial z} u'$
respectively) is the advection of the fluctuating velocity field by
the mean field, with the mean wall-normal velocity $\langle v \rangle$
being zero in TCF (plane-averaging the continuity equation and
integrating in the wall-normal direction from the wall leads to
$\langle v \rangle =0$); $u_j' u_{x_j}'$ (with Einstein summation over
$j=1,2,3$ and $u_{x_j}'$ standing for $\frac{\partial}{\partial x_{j}}
u'$) is the self-advection of fluctuating velocity field; $u_j'
\langle u_{x_j} \rangle = v'\langle u_{y} \rangle$ ($u_{x_j}$ stands
for $\frac{\partial}{\partial x_{j}} u$) is a term related to
turbulent kinetic energy production, as it involves the interaction
between the fluctuating velocity field and the mean shear; and finally
$\langle u_j' u_{x_j}' \rangle$ is the plane-average Reynolds shear
stress gradient, who does not correlate with fluctuating shear and
therefore has no contribution in the fluctuating wall shear stress.
The premultiplied covariances between the fluctuating streamwise shear
and each one of the four fluctuating components making up $a_u'$ in
equation (\ref{eq:audecomp}) are shown in figure \ref{fig:advec}.  The
sum of these four covariances is equal to the fluctuating
shear-acceleration covariance in figure \ref{fig:integrand}($a$).

Upon first observation, we notice that the magnitude of the covariance
of local shear with Eulerian acceleration (figure \ref{fig:advec}$a$)
and that with mean advection (figure \ref{fig:advec}$b$) are opposite
in sign and at least one order of magnitude larger than the overall
covariance between shear and Lagrangian acceleration in figure
\ref{fig:integrand}($a$), as if they almost cancel each other. It
%therefore makes sense
is therefore sensible to investigate whether and to what extent the
local Eulerian acceleration of the fluctuating streamwise velocity is
mainly due to local mean advection of that velocity, i.e. whether and
to what extent Taylor's frozen eddies hypothesis, $u_t' \approx -
\langle u_j \rangle u_{x_j}'$, holds.

\begin{figure}
\centering
\includegraphics[scale=1]{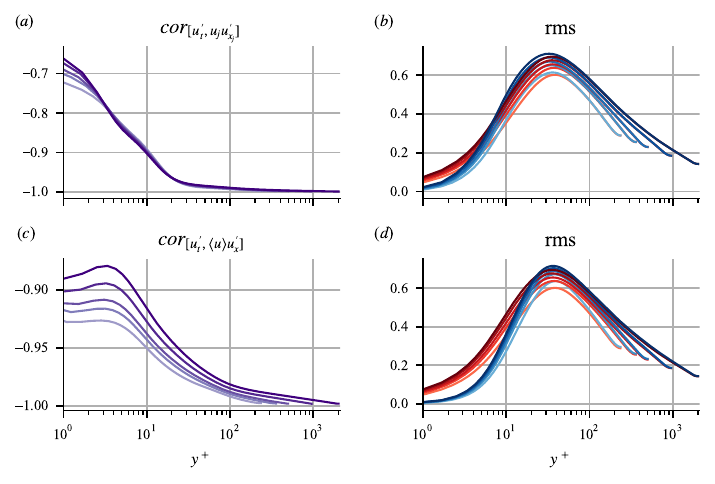}
\caption{Taylor's hypothesis tested with correlation coefficient and
  rms values. ($a$) and ($c$): correlation of Eulerian acceletation
  and streamwise advection. In ($b$) and ($d$), red: rms of
  fluctuating Eulerian acceleration. In ($b$) blue, rms of advection
  of fluctuating velocity by total velocity. In ($d$) blue, rms of
  mean advection of fluctuating velocity. All rms are normalised in
  inner units.}
\label{fig:utux}
\end{figure}

In figure \ref{fig:utux}($a$) we plot the spatial correlation
coefficient between Eulerian acceleration $u_t'$ and sum of streamwise
mean and self-advection $u_ju_{x_j}' = \langle u_j\rangle u_{x_j}'
+u_j' u_{x_j}' $, and in figure \ref{fig:utux}($b$) their respective
root mean square (rms) values.  Large negative correlation
coefficients, reaching
nearly -1 at channel centre, between the Eulerian acceleration and
streamwise advection across the channel indicate strong
anti-correlating behaviour far enough from the wall. This cancelation
behaviour is further consolidated by their similar rms values far
enough from the wall, e.g. $y^+ \gtrsim 50$. In fact, the
anti-correlation between the local Eulerian acceleration and the
streamwise mean advection $\langle u \rangle u_x'$ is even stronger,
as shown by the similar correlation coefficients in figure
\ref{fig:utux}($c$) and the rms peaks in figure \ref{fig:utux}($d$),
evidencing
a large degree of applicability of Taylor's hypothesis where this
anti-correlation is large enough. The spanwise mean advection is very
small due to the small plane-average spanwise velocity $\langle
w\rangle$, even in the current small computational domain.  Removing
the self-advection of fluctuating velocity field
leads to stronger anti-correlation close to the wall.  While the
turbulent self-advection is relatively small compared to the mean
advection, they still provide significant contribution to the overall
shear-acceleration covariance, as shown by the magnitude of the
covariance in figure \ref{fig:advec}($c$) compared to that in figure
\ref{fig:integrand}($a$).

It is noteworthy that Taylor's frozen eddies hypothesis appears valid
with correlation coefficients between $-0.85$ and $-1$ in figure
\ref{fig:utux}($c$) at all distances from the wall, which confirms
previous correlation analyses by \citet{piomelli_balint_wallace_1989}
and \citet{geng_he_wang_xu_lozanoduran_wallace_2015}. We can therefore
conclude that the average convective nature of TCF testified by
Taylor's hypothesis is largely responsible for the cancellation
between the covariance of local fluctuating shear with fluctuating
Eulerian acceleration (figure \ref{fig:advec}$a$) and the covariance
of local fluctuating shear with mean advection (figure
\ref{fig:advec}$b$) in this region. 
In the definition of the correlation coefficient (\ref{eq:cor}), the plane-average mean velocity appearing in the numerator is factored out by itself in the denominator. The choice of such plane-average velocity plays no role in the value of the coefficient.
The mean streamwise velocity $\langle u \rangle$ may
in fact differ from the actual convection velocity, especially in
the near-wall region, as detailed by
\citet{geng_he_wang_xu_lozanoduran_wallace_2015}.  As far as
covariances of plane-fluctuating quantities are concerned, the
choice of mean or convection velocity plays no role in the
correlation analysis.

We have already noted that the shear-acceleration covariance in figure
\ref{fig:integrand}($a$) provides the largest contribution to
$\overline{\langle {\tau_w'}^2\rangle }$. It has a peak close to about
$y^+ \approx 5$. Among the four covariances in figure \ref{fig:advec},
only the covariance between local fluctuating shear and the turbulence
production-related term displays a peak below $y^+<10$ (figure
\ref{fig:advec}$d$). It is this peak which is reflected in the
shear-acceleration covariance peak seen in figure
\ref{fig:integrand}($a$). These two peaks do not only share a similar
$y^+$ location, they also display similar peak magnitudes, although
the shear-acceleration covariance displays a Reynolds number
dependence that is more pronounced than the shear-production
covariance. We propose that it is the shear-production term that
mainly determines the near-wall peak position of the
shear-acceleration covariance in figure \ref{fig:integrand}($a$) and
principally contributes to its peak magnitude, and that the other
covariances in figure \ref{fig:advec} contribute to the Reynolds
number dependence.

\subsection{Origin of the near-wall peak}
\begin{figure}
\centering
\includegraphics[scale=1]{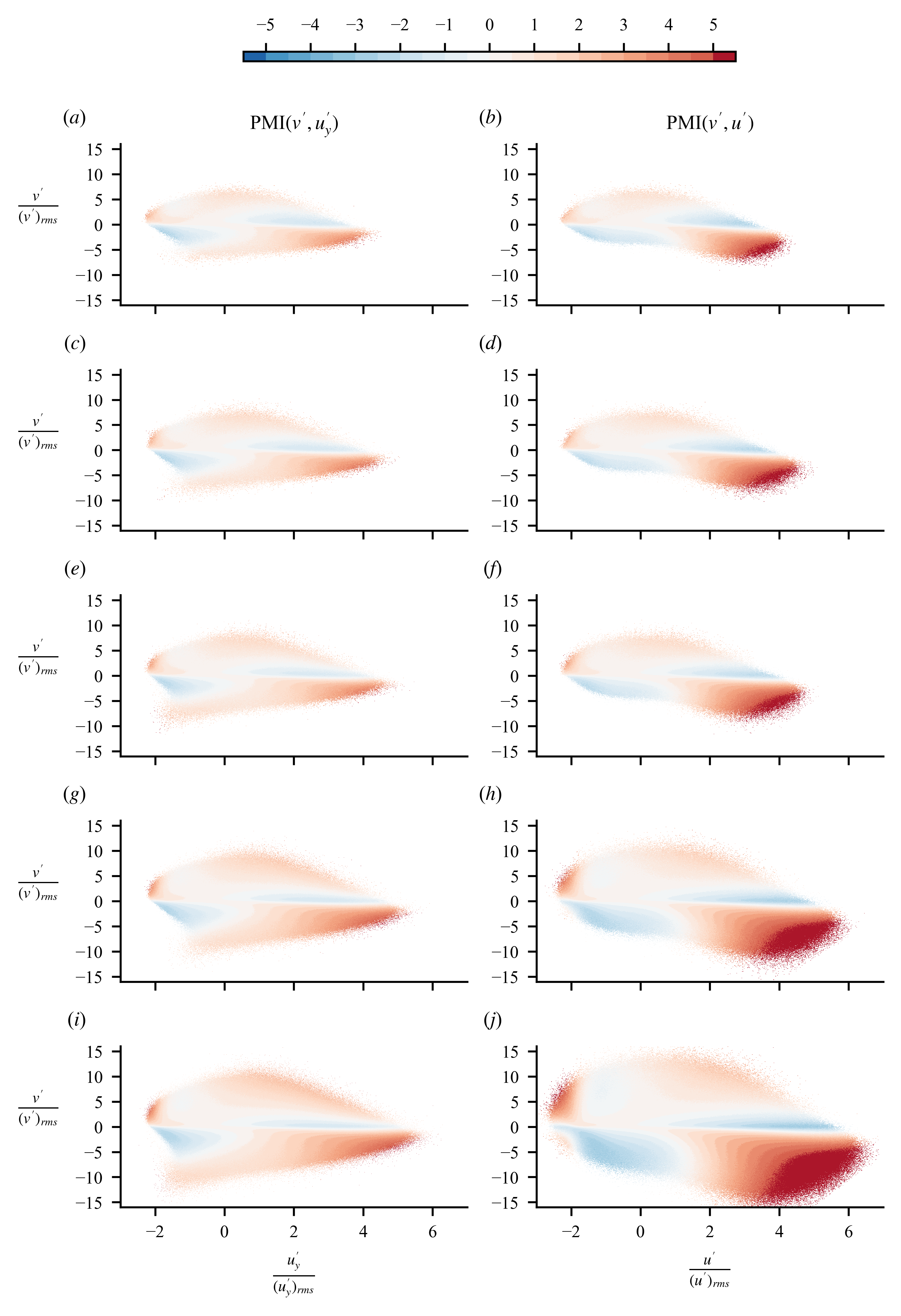}
\caption{Point-wise mutual information at $y^+\in [4,5]$ where the
  shear-production covariance $-\overline{\langle u_y\rangle\langle v'
    u_y'\rangle}$ maximises. Left: $(v',u_y')$; right:
  $(v',u')$. ($a,b$) R230, ($c,d$) R360, ($e,f$) R500, ($g,h$) R950,
  ($i,j$) R2000. All quantities are normalised by their respective rms
  values.}
\label{fig:MI}
\end{figure} 

In TCF, $\langle v \rangle(y,t) = 0$ at all $y$.  Inside the $x$-$z$
plane-average of shear-production covariance $-\langle
u_y\rangle\langle u_y'v' \rangle$, one may consider the fluctuating
local shear as a weight to the averaging of $v'$. For the weighted
average $-\langle u_y'v' \rangle$ to be positive within the viscous
sublayer, there must, for example, be an increased chance for $u_y'$
to be large and positive wherever $v'$ is negative, i.e. the two
quantities must be dependent in their joint probability density
function (JPDF) $P(v',u_y')$. If they were independent from each
other, we would have $P(v',u_y') = P(v')P(u_y')$ and then converting
plane-averaging into expectations on the support of $v'$ and $u_y'$:
\begin{equation*}
-\langle v' u_y' \rangle = \int P(v',u_y')~\mathrm{d}(v',u_y') =\int P(v')P(u_y')~\mathrm{d}v'\mathrm{d}u_y' = -\langle v' \rangle \langle u_y' \rangle = 0.
\end{equation*}

To demonstrate statistical dependence between two variables,
we compare directly the joint PDF to the marginal PDFs in terms of
point-wise mutual information (PMI) defined as:
\begin{equation}
\mathrm{PMI}(v',u_y') \equiv \ln\left( \frac{P(v',u_y')}{P(v')P(u_y')} \right) = \ln\left( \frac{P(v'|u_y')}{P(v')} \right) = \ln\left( \frac{P(u_y'|v')}{P(u_y')} \right)
\label{eq:MI}
\end{equation}
Given a set of values $(v',u_y')$, if the PMI is positive, $P(v',u_y')
> P(v')P(u_y')$ and the probability density of the given values is
greater than if the two variables were independent, hence they are
statistically correlated.  One may also interpret the PMI from a
Bayesian point of view: the positive value of PMI indicates that the
conditional probability of $v'$ provided a value of $u_y'$ is more
probable than the marginal probability of $v'$ without such condition
and vice versa.

We plot in figure \ref{fig:MI}($a,c,e,g,i$) the PMI of the fluctuating
wall-normal velocity and the fluctuating streamwise shear around
$y^+\in [4,5]$ where $-\overline{\langle u_y\rangle\langle v'
  u_y'\rangle}$ reaches its maximum in the viscous sublayer for five
Reynolds numbers.  A strong correlation with positive PMI is shown
between the negative values of $v'<0$ and large positive values of
$u_y'>0$, which contribute to the positive peak in $-\overline{\langle
  u_y\rangle\langle v' u_y'\rangle}$. To further elucidate the origin
of this local statistical dependence, we plot in figure
\ref{fig:MI}($b,d,f,h,j$) the pointwise mutual information for
$(v',u')$ at the same wall normal position.  This time, the negative
local $v'$ is only significantly correlated with large positive $u'$
(sweeps $u'>0,~v'<0$) but not negative $u'$ values. It is known that
in wall-bounded turbulence ejections $(u'<0,~v'<0)$ and sweeps
$(u'>0,~v'<0)$ make a dominant contribution to the Reynolds shear
stress \citep[see e.g.][]{pope_2000}. Here we find that sweeps
dominate the near-wall dynamics in the viscous sublayer with very high
statistical dependence in terms of PMI between large positive $u'$ and
negative $v'$. In the viscous sublayer very near the wall, high values
of $u'$ imply high values of $u_y'$, hence the PMI of $(v', u_y')$ may
be a consequence of the PMI of $(v', u')$ at sweep events. Ejections
also contribute to $-\overline{\langle u_y\rangle\langle v'
  u_y'\rangle}$ and its sign but less significantly than sweeps as
seen in both PMIs. Negative values of $u'$ imply negative $u_y'$ in
the viscous sublayer very near the wall and so ejections contribute to
the positive value $-\overline{\langle u_y\rangle\langle v'
  u_y'\rangle}$ in the viscous sublayer. In conclusion, sweeps, 
predominantly, but also ejections dominate the shear-production
covariance $-\overline{\langle u_y\rangle\langle v' u_y'\rangle}$
around $y^+\in [4,5]$ in the viscous sublayer and, in turn, the
near-wall peak of the shear-acceleration covariance in figure
\ref{fig:integrand}($a$). 

Sweeps typically originate far away from the wall so that their size
and strength should be, at least partially, characterised by outer
units at sufficiently high Reynolds number. They can therefore impose
some outer flow and Reynolds number dependence on the
shear-acceleration covariance and, as a result, on the variance of the
fluctuating wall shear stress. Furthermore, with increasing Reynolds
number, it is observed that the support of $u_y$ expands in figure
\ref{fig:MI}($a,c,e,g,i$), favouring increasing extreme events in the
near-wall region. As the fluctuating velocities and their gradients
become more intermittent and more extreme, the higher order statistics
will also start to show differences with increasing Reynolds number.
We observe that the second-order statistical quantity
$-\overline{\langle u_y\rangle\langle v' u_y'\rangle}$ is very weakly
dependent on Reynolds number below $y^+<10$ for our range of Reynolds
numbers.  On the contrary, the shear-self-advection covariance
$-\overline{\langle u_y' (u_j' u_{x_j}')\rangle}$ as a third-order
statistical quantity shows stronger variation with Reynolds number,
which might be explained by increased intermittency in the region not
very far from the wall (e.g. $y^+ \approx 10$).

%%%%%%%%%%%%%%%%%%%%%%%%%%%%%%%%%%%%%%%%%%%%%%%%%%%%%%%%%%%%%%%%%%%%%%
\section{Conclusion}
We have studied the relation between the wall shear stress and the
fluid flow above the wall in a fully developed turbulent channel flow
(FD TCF). Without any preconception (concerning, for example, energy
or momentum balance), the Navier-Stokes equations naturally imply that
the mean wall shear stress can be obtained from mean streamwise shear
and mean streamwise acceleration in the flow, and that the variance of
the fluctuating wall shear stress can be obtained from correlations
between fluctuating streamwise shear and fluctuating streamwise
acceleration predominantly as well as correlations between fluctuating
streamwise shear and fluctuating pressure gradient in the flow. At the
Reynolds numbers accessible by our FD TCF DNS data, it is enough to
integrate these flow statistics from the wall to the outer edge of the
buffer layer, see equations (\ref{eq:pavgapprox}) and
(\ref{eq:flucapprox}). In other words, information within the buffer
layer is sufficient to determine the mean wall shear stress and the
variance of the fluctuating wall shear stress to very good accuracy,
at least at the Reynolds numbers considered here. Above the buffer
layer, the fluctuating streamwise acceleration near-balances the
fluctuating pressure gradient so that the correlations between
fluctuating streamwise shear and fluctuating streamwise acceleration
and the correlations between fluctuating streamwise shear and
fluctuating pressure gradient effectively cancel each other. The outer
bound $\delta_{w}$ in the integrals in these two equations appears to
scale with the wall unit $\delta_{\nu}$, but very weak dependencies of
$\delta_w$ on $Re_{\tau}$ are conceivable in equations
(\ref{eq:pavgapprox}) and (\ref{eq:flucapprox}). In fact there may be
two different $\delta_w$ scales defined by slightly different Reynolds
number dependencies, a weaker one for (\ref{eq:pavgapprox}) and a
slightly stronger one for in (\ref{eq:flucapprox}). Our data do not
reach high enough values of $Re_{\tau}$ to determine these $Re_{\tau}$
dependencies.

Even though equations (\ref{eq:pavgapprox}) and (\ref{eq:flucapprox})
imply that one can find between the wall and the outer edge of the
buffer layer all the flow information required for a good estimation
of the mean wall shear stress and the variance of the wall shear
stress fluctuations, this does not mean that the wall shear stress
statistics do not depend on the outer flow. The mean streamwise
acceleration, the correlation between fluctuating streamwise shear and
fluctuating streamwise acceleration and the correlation between
fluctuating streamwise shear and fluctuating pressure gradient between
the wall and the buffer layer have Reynolds number dependencies which
betray a dependence on the outer flow.

Some understanding of the correlations between fluctuating streamwise
shear and fluctuating streamwise acceleration which dominate the
variance of the fluctuating wall shear stress has been obtained by
using a Reynolds decomposition of the streamwise velocity to decompose
the fluctuating streamwise acceleration into four terms plus a plane-average term (see equation
\ref{eq:audecomp}). This approach reveals cancellations caused by
Taylor's frozen eddies hypothesis which proves to be a good
approximation at all distances from the wall.

Finally, it is the turbulence production-related term in the
decomposition (\ref{eq:audecomp}) of the fluctuating streamwise
acceleration which accounts for the near-wall peak of the
shear-acceleration covariance in the viscous sub-layer. Sweeps, and to
a lesser extent, ejections dominate the shear-production covariance
$-\overline{\langle u_y\rangle\langle v' u_y'\rangle}$ around $y^+\in
[4,5]$ and, in turn, the near-wall peak of the shear-acceleration
covariance in the viscous sub-layer.

%%%%%%%%%%%%%%%%%%%%%%%%%%%%%%%%%%%%%%%%%%%%%%%%%%%%%%%%%%%%%%%%%%%%%%
\backsection[Funding]{This work is supported by the European Office of Aerospace Research and Development (EOARD) (FA8655-21-1-7016; Program Manager: Dr D. Smith).}

\backsection[Acknowledgements]{This work is granted access to the HPC resources of IDRIS under the allocation 2022-021741 made by GENCI (Grand Equipement National de Calcul Intensif) and Zeus supercomputers (Mésocentre de Calcul Scientifique Intensif de l’Université de Lille)}

\backsection[Declaration of interests]{The authors report no conflict of interest.}

%%%%%%%%%%%%%%%%%%%%%%%%%%%%%%%%%%%%%%%%%%%%%%%%%%%%%%%%%%%%%%%%%%%%%%
\appendix
\section{Derivation of $n$th order wall shear stress equation}\label{sec:A_nth}
The plane-fluctuating streamwise Navier-Stokes equation is:
\begin{equation}
a_u' = -p_x' + \nu u_{//}'+ \nu u_{yy}' .
\end{equation}
We multiply the fluctuating equation with $n (\nu u_y')^{n-1}$ at the same $(x,y,z)$ for some integer $n>1$, the wall-normal diffusion term (due to product rule) becomes:
\begin{equation}
n (\nu u_y')^{n-1} \nu u_{yy}'  = \frac{\partial}{\partial y} (\nu u_y')^{n}.
\label{eq:productrule}
\end{equation}

Averaging the premultiplied fluctuating streamwise Navier-Stokes
equation in $x$-$z$ plane and integrating from the wall up to some
arbitrary $y$:

\begin{eqnarray}
\int_0^y \langle n (\nu u_y')^{n-1} a_u'\rangle ~\mathrm{d}y  &= & -\int_0^y \langle n (\nu u_y')^{n-1} p_x'\rangle ~\mathrm{d}y \nonumber \\
&&+ \int_0^y \langle n (\nu u_y')^{n-1} \nu u_{//}'\rangle ~\mathrm{d}y + \int_0^y \langle n (\nu u_y')^{n-1} \nu u_{yy}'\rangle ~\mathrm{d}y \nonumber \\
&\Downarrow & \mathrm{using~} (\ref{eq:productrule})\nonumber \\
\int_0^y \langle n (\nu u_y')^{n-1} a_u'\rangle ~\mathrm{d}y  &= & -\int_0^y \langle n (\nu u_y')^{n-1} p_x'\rangle ~\mathrm{d}y \nonumber \\
&&+ \int_0^y \langle n (\nu u_y')^{n-1} \nu u_{//}'\rangle ~\mathrm{d}y +\langle (\nu u_y')^{n} \rangle(y) - \langle (\nu u_y')^{n}\rangle(y=0)  \nonumber
\end{eqnarray}

Rearranging, we obtain an equation for $n$th order wall shear stress:
\begin{align}
\langle {\tau_w'}^n \rangle 
= &  - n\nu^{n-1} \int_0^y \langle {u_y'}^{n-1} a_u'\rangle ~\mathrm{d}y \nonumber\\
&-n\nu^{n-1}\int_0^y \langle {u_y'}^{n-1} p_x'\rangle ~\mathrm{d}y \nonumber \\ 
& + n\nu^{n}\int_0^y \langle {u_y'}^{n-1}  u_{//}'\rangle ~\mathrm{d}y \nonumber\\
&+\langle (\nu u_y')^{n} \rangle(y)
\end{align}

%%%%%%%%%%%%%%%%%%%%%%%%%%%%%%%%%%%%%%%%%%%%%%%%%%%%%%%%%%%%%%%%%%%%%%
\bibliographystyle{jfm} \bibliography{jfm}
\end{document}